\documentclass[sigconf]{acmart}

\usepackage{graphicx}
\usepackage{epstopdf}
\usepackage{multirow}
\usepackage{balance}
\usepackage[caption=false,font=footnotesize]{subfig}
\usepackage{stfloats}
\usepackage{algorithmicx}
\usepackage{algpseudocode}
\usepackage{amsfonts}
\usepackage[linesnumbered,ruled,lined]{algorithm2e}
\usepackage{tabu}
\usepackage{upgreek}

\acmYear{2023}\copyrightyear{2023}
\acmConference[ISACom '23]{3rd ACM MobiCom Workshop on Integrated Sensing and Communication Systems for IoT (ISACom)}{October 6, 2023}{Madrid, Spain}
\acmBooktitle{3rd ACM MobiCom Workshop on Integrated Sensing and Communication Systems for IoT (ISACom) (ISACom '23), October 6, 2023, Madrid, Spain}
\acmPrice{15.00}
\acmDOI{10.1145/3615984.3616502}
\acmISBN{979-8-4007-0364-5/23/10}

\begin{document}

\title[5G NR Monostatic Positioning with Array Impairments: Data-and-Model-Driven Framework and Experiment Results]{5G NR Monostatic Positioning with Array Impairments: Data-and-Model-Driven Framework and Experiment Results}

\author{Shengheng Liu}
\authornote{Both authors are also affiliated to the Purple Mountain Laboratories, Nanjing 211111, China.}
\affiliation{%
  \institution{Southeast University School of Information Science and Engineering}
  \city{Nanjing 210096}
  \country{China}
}
\email{s.liu@seu.edu.cn}

\author{Hao Wang}
\affiliation{%
  \institution{ChuHang Technology Co. Ltd.}
  \city{Nanjing 211500}
  \country{China}
}

\author{Mengguan Pan}
\author{Peng Liu}
\affiliation{%
 \institution{Purple Mountain Laboratories}
 \city{Nanjing 211111}
 \country{China}}

\author{Yahui Ma}
\affiliation{%
  \institution{China Academy of Electronics and Information Technology}
  \city{Beijing 100041}
  \country{China}}

\author{Yongming Huang}
\authornotemark[1]
\affiliation{%
  \institution{Southeast University School of Information Science and Engineering}
  \city{Nanjing 210096}
  \country{China}}

\renewcommand{\shortauthors}{Liu et al.}

\begin{abstract}
In this article, we present an intelligent framework for 5G new radio (NR) indoor positioning under a monostatic configuration. The primary objective is to estimate both the angle of arrival and time of arrival simultaneously. This requires capturing the pertinent information from both the antenna and subcarrier dimensions of the receive signals. To tackle the challenges posed by the intricacy of the high-dimensional information matrix, coupled with the impact of irregular array errors, we design a deep learning scheme. Recognizing that the phase difference between any two subcarriers and antennas encodes spatial information of the target, we contend that the transformer network is better suited for this problem compared to the convolutional neural network which excels in local feature extraction. To further enhance the network's fitting capability, we integrate the transformer with a model-based multiple-signal-classification (MUSIC) region decision mechanism. Numerical results and field tests demonstrate the effectiveness of the proposed framework in accurately calibrating the irregular angle-dependent array error and improving positioning accuracy.
\end{abstract}



\begin{teaserfigure}
  \includegraphics[width=\textwidth,trim=0 360 0 1300,clip]{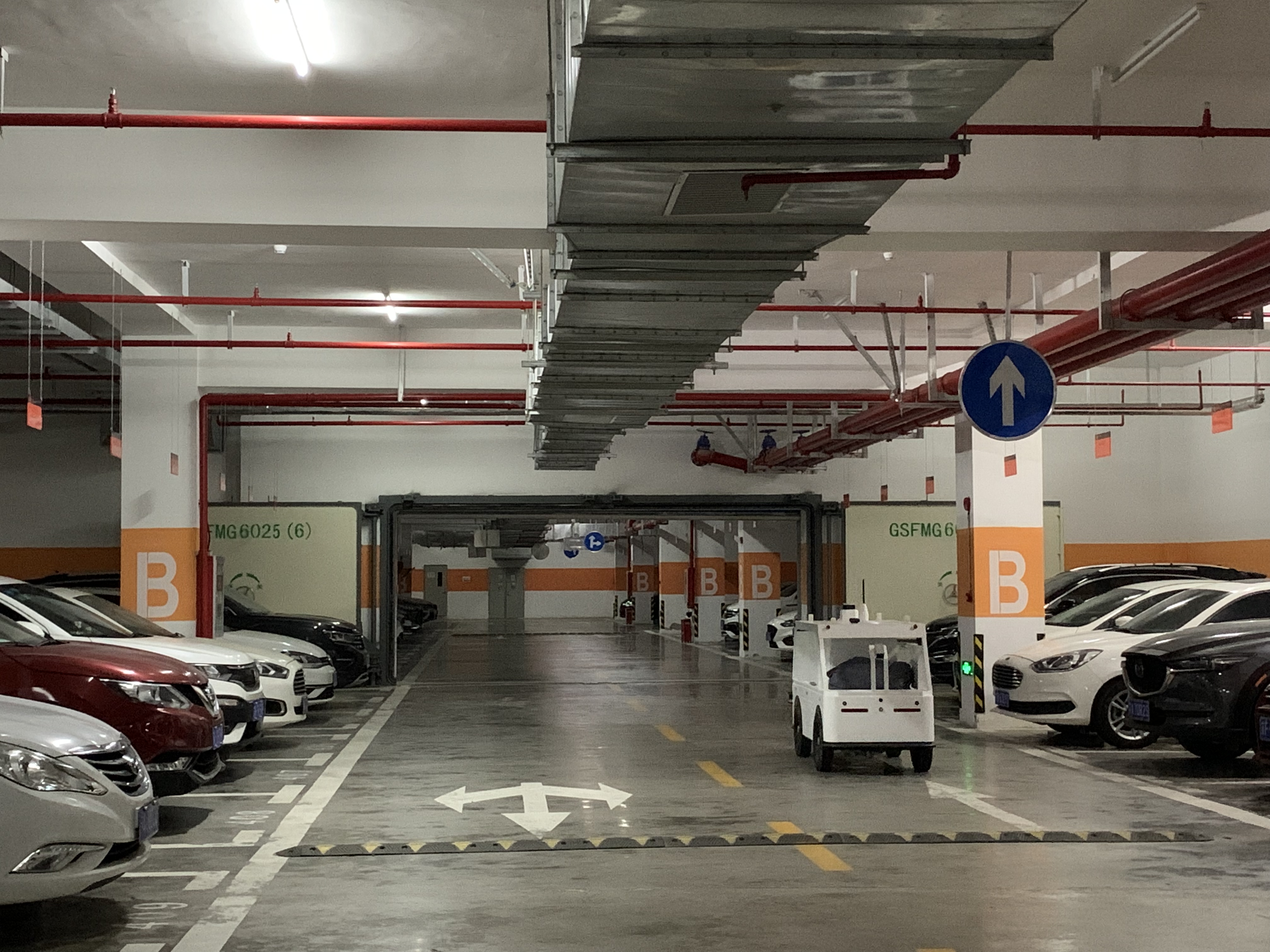}
  \caption{Automated guided vehicle (AGV) is automatically measuring and recording data from various onboard sensors, including the 5G terminal, in the underground parking lot of Purple Mountain Laboratories.}
  \label{fig:teaser}
\end{teaserfigure}

\received{18 June 2023}
\received[accepted]{31 July 2023}

\maketitle

\section{Introduction}

In the era of the internet of everything (IoE), the importance of location awareness spans across a broad spectrum of applications, such as autonomous driving, emergency relief, and assisted living. While global positioning system (GPS) stands as the dominant technology for outdoor environments, furnishing robust and precise positioning, its efficacy is hindered in deep urban canyons and indoor settings due to the limited power and weak signal penetration. Consequently, there exists a pressing imperative to develop innovative methodologies that facilitate accurate and reliable positioning in GPS-constrained environments. Such methodologies must deliver robust and precise location information, particularly in scenarios where GPS signals are attenuated or entirely unavailable. Moreover, an additional objective should be to minimize reliance on the procurement and installation of extra equipment, instead emphasizing the utilization of existing infrastructure to the greatest extent feasible, thereby mitigating deployment costs.

In the pursuit of advancing location-awareness techniques, researchers are driven to explore alternative strategies that effectively leverage available resources. The ubiquity of 5G signals in diverse indoor environments \cite{You23}, accompanied by their large bandwidth and received power, alongside the incorporation of antenna array technology within the transceiver, presents a compelling avenue for augmenting the performance of localization systems \cite{Groschel17}. Nevertheless, despite these promising attributes, significant challenges persist in attaining high-accuracy positioning, stemming from factors such as unknown hardware impairments and the synthesis errors arising from their intricate combination. The effective resolution of these challenges assumes utmost importance in fully unleashing the immense potential of 5G technology in the realm of location based services (LBS).

Data-driven approaches based on neural networks have gained significant popularity in the domain of localization algorithm design. Fingerprint-based positioning has emerged as a popular methodology \cite{Wang23}, characterized by the establishment of an offline fingerprint database and subsequent online data-driven similarity analysis of fingerprint patterns. Nonetheless, the efficacy of this approach is reliant on meticulous environmental measurements and is limited by its incapacity to adapt to dynamically evolving environments. Alternative algorithms have been proposed that leverage the estimated covariance matrix elements of the received signals as inputs, resulting in streamlined structures, expedited computation, and the establishment of a fundamental mapping relationship facilitated by fully connected layers \cite{Papageorgiou21}. Nevertheless, these algorithms often encounter limitations in their ability to estimate angular parameters and neglect the valuable subcarrier information embedded within 5G signals, consequently encountering challenges in monostatic localization. To address this issue, An et al. \cite{wanghao_60} proposed an innovative solution by employing spatial spectrums as inputs. This approach effectively extracts spatial dependence from the spatial spectrum, capitalizing on the capabilities of convolutional neural networks (CNNs) commonly utilized in image processing. Consequently, this method enables more accurate localization, albeit at the expense of soaring complexity during the generation of the spatial spectrum. In a separate study, Jang et al. \cite{wanghao_59} took a different approach by utilizing the signal's time-frequency responses as inputs. However, this methodology remained inherently one-dimensional in nature, failing to fully exploit the rich spatial phase information available, which is crucial for accurate localization.

This paper presents a novel deep learning-based algorithm for joint estimation of angle of arrival (AoA) and time of arrival (ToA) using a transformer network. To effectively utilize the phase differences across arbitrary antennas, the transformer network is specifically devised for comprehensive global information processing. In order to exploit the characteristics of phase errors observed in an anechoic chamber, the angular space is partitioned into two distinct regions: large angle and small angle regions. The training input consists of channel state information (CSI) derived from orthogonal frequency division multiplexing (OFDM) signals. To enhance the fitting performance in each region, two independent transformer networks are trained in parallel. Each network processes the CSI information and computes the correlations among the blocks. Given that the phase difference between any two blocks encapsulates the AoA and ToA information of the target, the transformer network emerges as a more suitable choice compared to CNNs for addressing this specific problem. To determine the final estimation results, a regional decision mechanism based on the multiple signal classification (MUSIC) is employed, enabling the selection from the outputs of the two Transformer networks. Through extensive experiments, we showcase that the proposed algorithm yields substantial enhancements in the accuracy of AoA and ToA estimation.

\section{Preliminary}

\subsection{Signal model}

The 3rd Generation Partnership Project (3GPP) introduced a range of new radio (NR) high-precision positioning technologies in the Release 16 version of the 5G standard \cite{3GPP_20}. Among these technologies, the upstream positioning scheme stands out as it solely necessitates the transmission of uplink-sounding reference signal (UL-SRS) from users to base stations for position determination at the base station side. This scheme offers enhanced accuracy and sensitivity to direction angles while maintaining a balance between positioning performance and user equipment (UE) requirements \cite{Abu-Shaban18}. Consequently, this paper focuses primarily on the upstream positioning pathway. In this context, we suppose that the UE transmits the SRS sequence, denoted as $\mathbf{s}(t)$, which occupies $M$ subcarriers. The UL-SRS is received by a uniform linear array comprising $N$ receive antennas deployed at a base station. For each discrete time instance $t=1, 2, \ldots, T$, the received signal vector, corrupted by noise $\mathbf{e}(t)$, can be expressed as
\begin{align}
	\mathbf{y}(t)=\mathbf{a}\left(\theta_d, \tau_d\right) \cdot\mathbf{s}(t)+\mathbf{e}(t),
	\label{f01}
\end{align}
where $\theta_d$ and $\tau_d$ are respectively the AoA and ToA of the UE. $\mathbf{a}\left(\theta_d, \tau_d\right)$ denotes the steering vector. In particular, the relative phase difference of the $m$-th element, with the first antenna as a reference, can be expressed as
\begin{align}
	\Phi_{\theta_d}^m=\exp \left\{-{\rm j} 2 \uppi d(m-1) \sin \theta_d / \lambda\right\},
	\label{f02}
\end{align}
where $d$ represents the inter-element spacing and $\lambda$ denotes the carrier wavelength. Similarly, with the first subcarrier as a reference, the relative phase difference of the $n$-th subcarrier is written as
\begin{align}
	\Psi_{\tau_d}^n=\exp \left\{-{\rm j} 2 \uppi(n-1) \cdot\Delta f \cdot\tau_d \right\},
	\label{f03}
\end{align}
where $\Delta f$ is the subcarrier spacing for the OFDM system. Combining the phase differences $\Phi_{\theta_d}$ and $\Psi_{\theta_d}$, the steering vector for the joint estimation problem of AoA and ToA can be derived as follows
\begin{align}
	\mathbf{a}\left(\theta_d, \tau_d\right)=\left[1, \ldots, \Psi_{\tau_d}^N, \ldots, \Psi_{\tau_d}^1 \Phi_{\theta_d}^M, \ldots, \Psi_{\tau_d}^N \Phi_{\theta_d}^M\right].
	\label{f04}
\end{align}
Taking into account the effect of directional hardware impairment, the received signal can be modelled as
\begin{align}
	\mathbf{x}(t)=\boldsymbol{\Gamma} \odot \mathbf{a}\left(\theta_d, \tau_d\right) \mathbf{s}(t)+\mathbf{e}(t),
	\label{f05}
\end{align}
where matrix $\boldsymbol{\Gamma}$ captures the phase error of each antenna. In the case where only gain and phase inconsistencies are considered, $\boldsymbol{\Gamma}$ can be described using a diagonal matrix, with the $l$-th diagonal entry being the coefficient of the $l$-th radio frequency (RF) channel.

\subsection{Analysis of phase characteristics using anechoic chamber measurement}

\begin{figure}[h]
\centering
\includegraphics[width=0.4\textwidth]{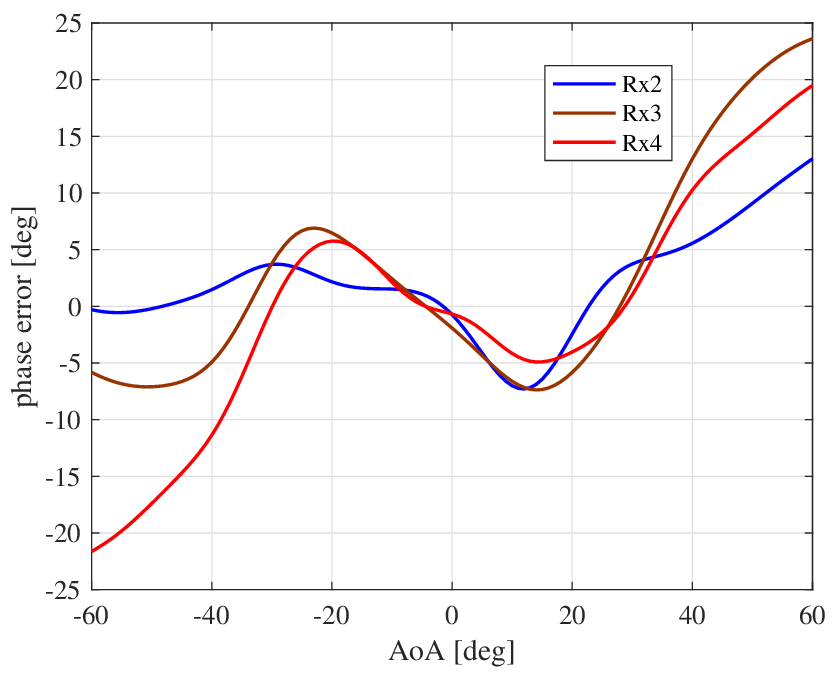}
\vspace{-1em}
\caption{Anechoic chamber measurements of a commercial sub-6GHz four-antenna 5G gNB operating at 2.565 GHz, which exhibits angle-dependent phase errors.}
\label{p05}
\end{figure}

To gain insights into the characteristics of the phase errors, we have previously conducted a series of measurement experiments in a microwave anechoic chamber. 
During these experiments, a commercial four-antenna 5G gNB was subjected to testing. By placing the array antenna on a movable turntable, we collected CSI data of each antenna in different directions of arrival. Analyses of the measurement data, as depicted in Fig.~\ref{p05}, revealed irregular anisotropic properties in the phase errors of the second to fourth antennas relative to the first antenna of the 5G gNB. It is important to note that such array errors, prevalent in real-world systems, can have a detrimental impact on the orthogonality between signal and noise subspaces. As a consequence, the performance of conventional purely model-driven spatial spectrum estimation methods can experience significant degradation.

\section{Methodology}

The accuracy of the commonly used MUSIC algorithm is considerably compromised due to the detrimental effects of array errors and multipath propagation. Furthermore, in the context of joint AoA-ToA estimation, the high-dimensional matrix eigenvalue decomposition and two-dimensional grid search entail substantial computational complexity. To tackle these challenges, we design a transformer network for feature extraction. By leveraging the distinctive characteristics of anisotropic array errors, we partition the angle space and employ parallel training to enhance the localization accuracy and robustness.

\subsection{Data preprocessing}

In most cases, the dimensionality of the received subcarrier information exceeds that of the antenna dimension. Therefore, when the received signal is fed to a neural network as a one-dimensional vector, it tends to restrict the network's capacity to extract features along the desired target arrival angle. Instead, the network tends to capture features primarily along the subcarrier axis. Thus, it becomes imperative to reorganize the ideal steering vector matrix accordingly, which can be expressed as follows.
\begin{align}
	\widetilde{\mathbf{A}}\left(\theta_d, \tau_d\right)=\left[\begin{array}{cccc}
		1 & \Psi_{\tau_d}^2 & \cdots & \Psi_2^N \\
		\Phi_{\theta_d}^2 & \Phi_{\theta_d}^2 \Psi_{\tau_d}^2 & \cdots & \Phi_{\theta_d}^2 \Psi_{\tau_d}^N \\
		\vdots & \vdots & \ddots & \vdots \\
		\Phi_{\theta_d}^M & \Phi_{\theta_d}^M \Psi_{\tau_d}^2 & \cdots & \Phi_{\theta_d}^M \Psi_{\tau_d}^N
	\end{array}\right].
	\label{f05071}
\end{align}
In this rearranged matrix, the phase differences between rows and columns are determined by the antenna array and subcarrier dimensions, respectively. Simultaneously, the single-snapshot received signal is also reshaped into a corresponding matrix form denoted as $\mathbf{Y} \in \mathbb{C}^{M \times N}$. Lastly, the received signal matrix $\mathbf{Y}$ is split into its real part $\mathbf{Y_1}$ and imaginary part $\mathbf{Y_2}$, facilitating the use of a dual-channel transformer neural network for subsequent processing.

\subsection{Training phase}

\begin{figure}[!t]
	\centering
	\includegraphics[width=0.45\textwidth]{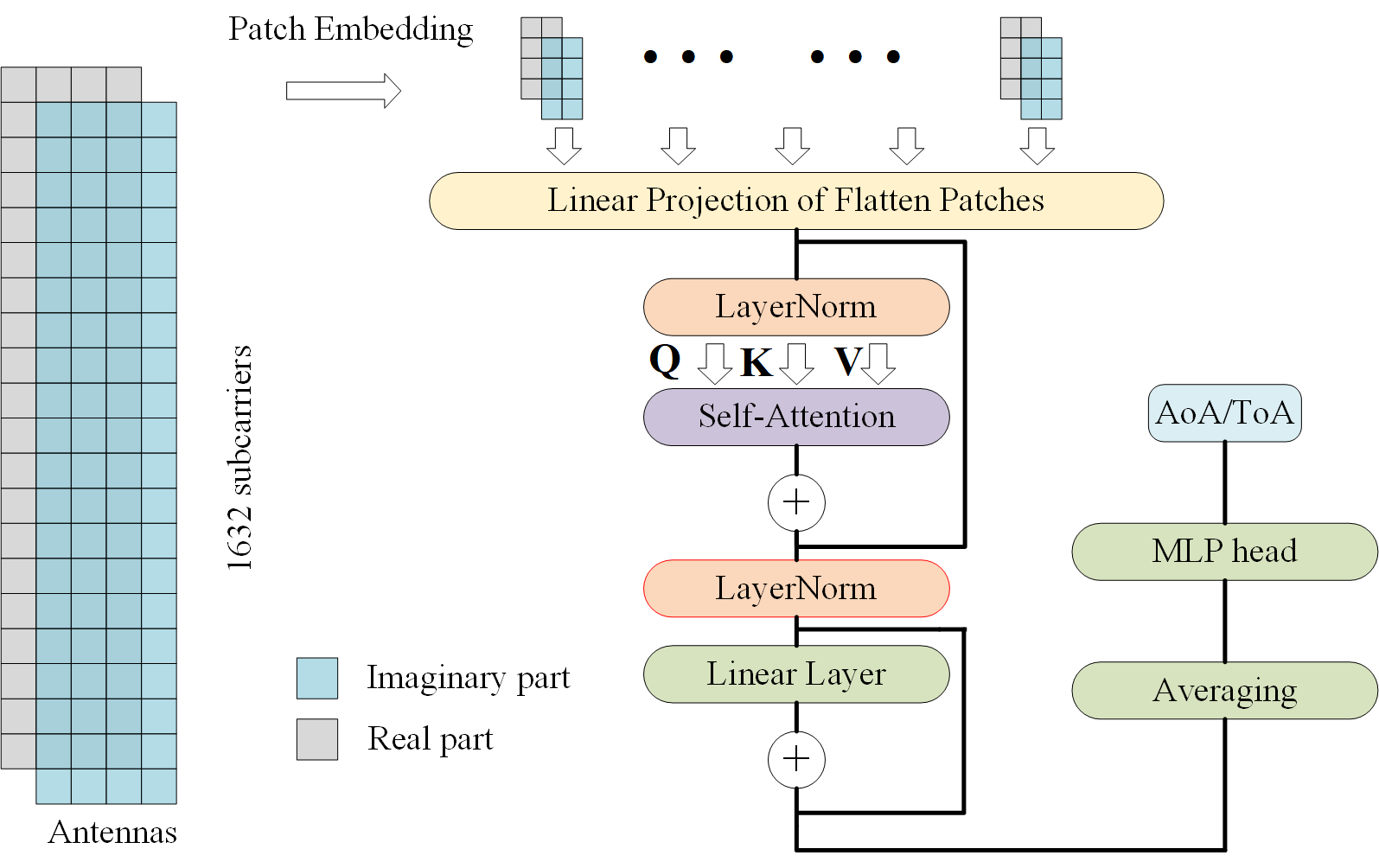}
\vspace{-1em}
	\caption{Architecture of the proposed transformer network for joint AoA-ToA estimation.}
	\label{p0509}
\end{figure}

The employed network architecture, as illustrated in Fig.~\ref{p0509}, encompasses several key components. Firstly, patch embedding is performed on the input CSI data. This process involves dividing the CSI into blocks, wherein each block matrix captures the phase relations between different antennas and subcarriers.

The block matrices undergo a reconfiguration process where their entries are transformed into a one-dimensional vector. This vector is then fed into the Linear Projection module in Fig.~\ref{p0509}, which is a weight-sharing, single-layer fully connected neural network. This module is responsible for extracting pertinent and informative features from the block matrix. The output vectors of linear projection module are subsequently directed to the LayerNorm layer. This layer serves the purpose of mitigating data saturation within the activation function, thereby addressing the issue of gradient vanishing. Following this, the vectors are concatenated to create a comprehensive output matrix. Once this output matrix is obtained, three weight matrices, namely the query matrix $\mathbf{Q}$, the key matrix $\mathbf{K}$, and the value matrix $\mathbf{V}$, are constructed and assigned to the Self-Attention module. The Self-Attention module manipulates these matrices in accordance with \eqref{f512} to generate the attention map.

\begin{align}
	\mathbf{A}=\operatorname{softmax}\left(\frac{\mathbf{Q} \mathbf{K}^{\top}}{\sqrt{d_k}}\right),
	\label{f512}
\end{align}
where $d_k$ represents the length of the output vector. The division by the square root of this value, as described in \eqref{f512}, serves the purpose of ensuring a more stable gradient during the training phase. Here, the softmax function is applied row-wise to the matrix, i.e.,
\begin{align}
	\operatorname{Softmax}\left(z_i\right)=\frac{\exp \left(z_i\right)}{\sum_j \exp \left(z_j\right)}.
	\label{f513}
\end{align}
Once the attention map is obtained, it is multiplied with the value matrix, yielding the output matrix.
\begin{align}
	\mathbf{Out}=\operatorname{softmax}\left(\frac{\mathbf{Q K}^{\top}}{\sqrt{d_k}}\right) \mathbf{V}.
	\label{f514}
\end{align}
This output matrix can subsequently be passed through one or multiple fully connected neural networks to extract features at deeper levels. To enhance the model's capability in fitting non-linear errors, we adopt a separate training strategy involving two independent transformers, with one specifically tailored for the large angular region $\left[ { - 60^\circ , - 45^\circ } \right) \cup \left( {45^\circ ,60^\circ } \right]$ and the other for the small angular region $\left[ { - 45^\circ ,45^\circ } \right]$.

\subsection{Validation phase}

\begin{figure*}[!t] \subfloat[]{\includegraphics[width=0.333\textwidth]{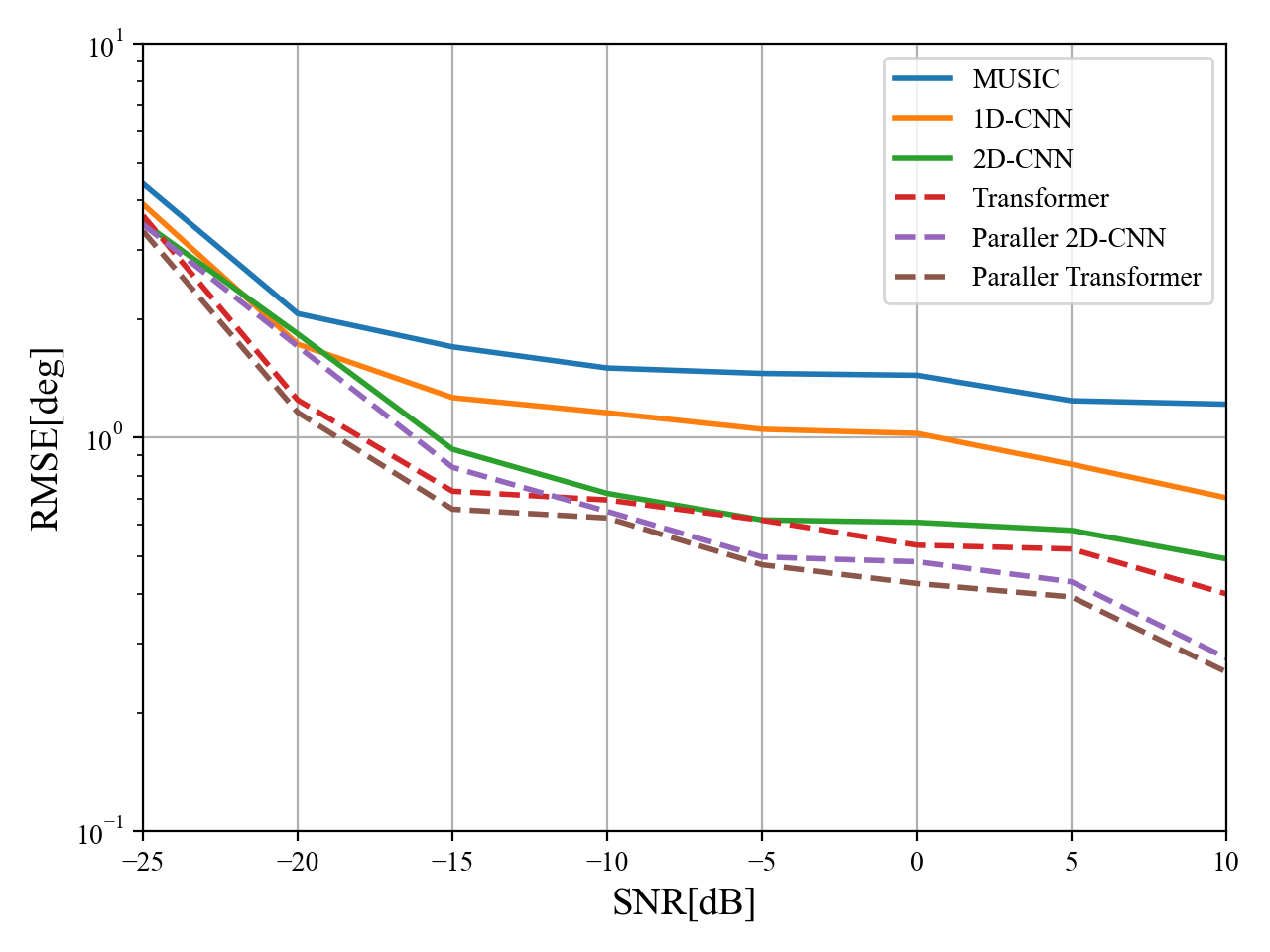}}	\subfloat[]{\includegraphics[width=0.333\textwidth]{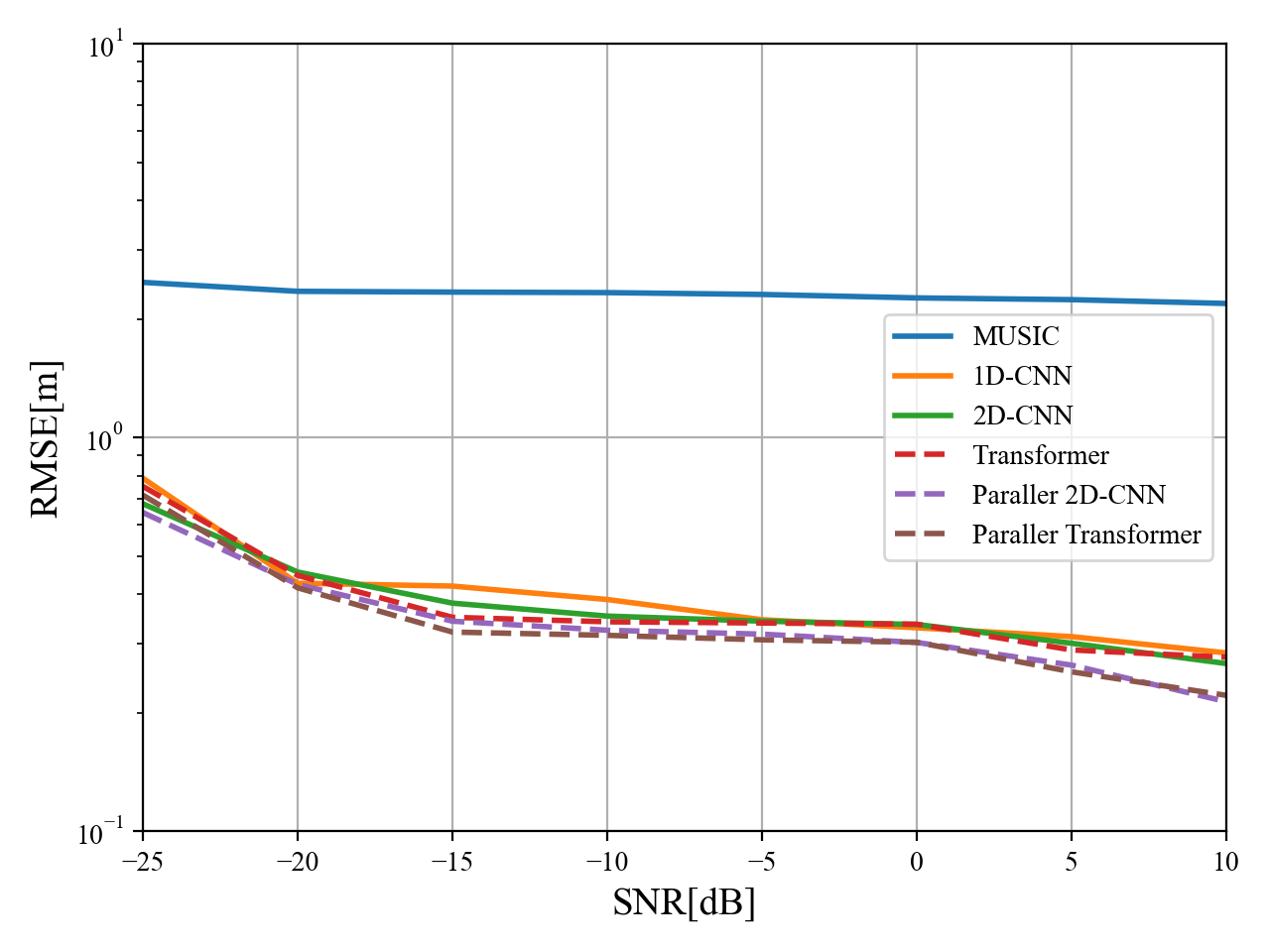}}		\subfloat[]{\includegraphics[width=0.333\textwidth]{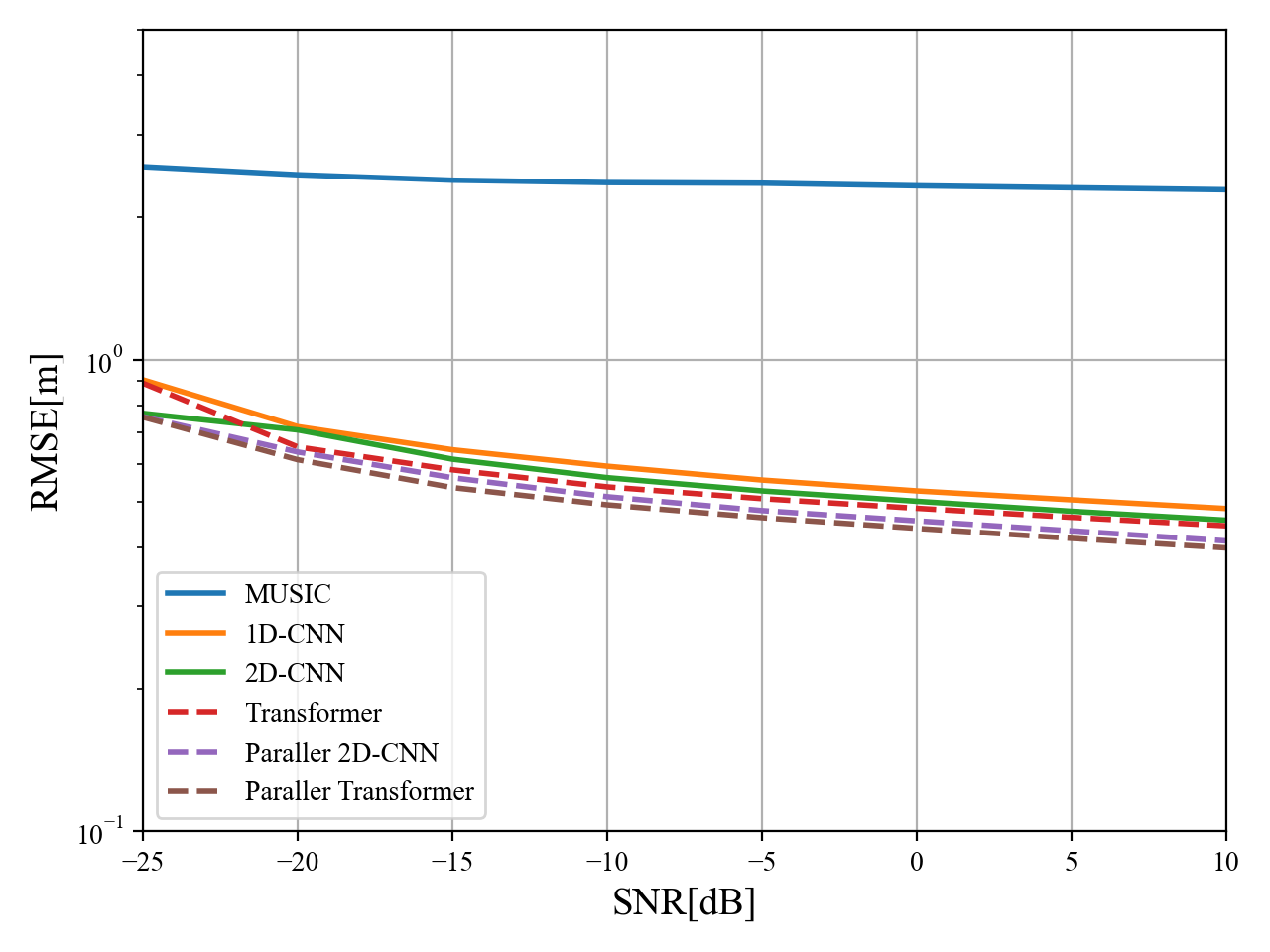}}
\vspace{-1em}
\caption{Performance comparison of different methods. (a) RMSE of AoA estimation versus SNR. (b) RMSE of ToA estimation versus SNR. (c)RMSE of localization estimation versus SNR. }
\label{p01}
\end{figure*}

In the training phase, we partition the angular region and train separate neural networks on the corresponding data segments. As a result, a decision criterion is required during the network validation stage to determine whether to employ the large-angle or small-angle region neural network for validation. For this purpose, we employ the MUSIC algorithm as the basis of decision mechanism in the proposed scheme. The selection of the MUSIC algorithm for validation is grounded in the observation that the nonlinearity arising from anisotropic deviations primarily manifests within the angular domain, while deviations within the frequency dimension are predominantly linear. As such, we regard the subcarrier dimension of the 5G signals as akin to a frequency snapshot. This implies that even when only a single snapshot signal is available in the time domain, it remains possible to accurately estimate the covariance matrix of the received signal. The covariance matrix is estimated by ${\bf{\hat R}} = {\bf{Y}}{{\bf{Y}}^{\mathsf{H}}}$. At this stage, a spectral peak search solely within the angle dimension is required, thereby significantly reducing the computational complexity. Although this introduces a certain level of decision error, it significantly enhances the network's fitting capability across various regions.

\section{Performance Evaluation}

This section focuses on the evaluation of the proposed scheme's performance through numerical simulation and real-world measurement data. The objective is to analyze the effectiveness of proposed data-and-model-driven scheme in comparison with several deep learning algorithms and traditional spatial spectrum estimation methods for monostatic positioning, particularly in the presence of array errors and multipath interference. In the investigation, we consider a pico-cell base station equipped with $N = 4$ array elements and $M = 1632$ subcarriers. The carrier frequency is set at $2.565\;{\rm{GHz}}$, with a subcarrier interval of $30\;{\rm{KHz}}$.

In the numerical simulations using synthetic data, we employed the link-level simulator that was previously presented in our prior work \cite{Jia23}. This simulator has been designed to accommodate a range of standard 5G scenarios, including indoor, indoor factory, urban macro, urban micro, and rural macro environments. The parameters associated with these scenarios adhere to the 3GPP TR 38.901 specification of wireless channel model. For the numerical simulations conducted in this study, we specifically selected a typical indoor scene that aligns with the settings of our field test. During generation of the training data, we vary the signal-to-noise ratio (SNR) from $-10\;{\rm{dB}}$ to $30\;{\rm{dB}}$, the target distance from $3\;{\rm{m}}$ to $10\;{\rm{m}}$, and the target angle from $-60^\circ$ to $60^\circ$. These parameter variations allow us to comprehensively evaluate the algorithm's performance under diverse conditions. We generated a dataset comprising a total of $20,000$ samples through simulation. The data was divided into two subsets with $\pm$$45^\circ$ as boundaries, and two networks were trained in parallel.

\begin{figure}[h]
\centering
\includegraphics[width=0.38\textwidth]{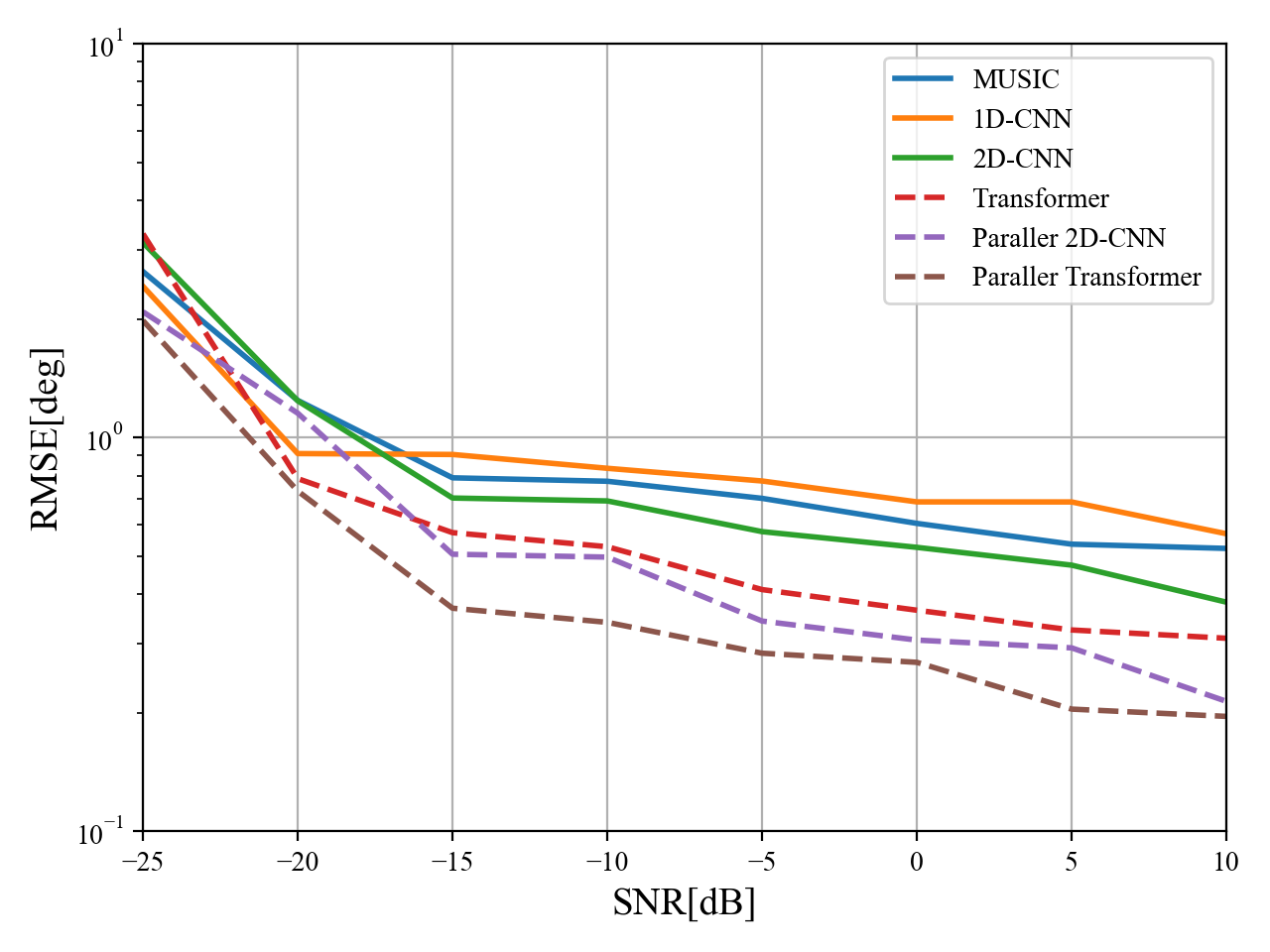}
\vspace{-1.5em}
\caption{RMSE of AoA estimation in small angle region versus SNR.}
\label{pic2}
\end{figure}

\begin{figure}[!htpb]
\centering
\includegraphics[width=0.49\textwidth]{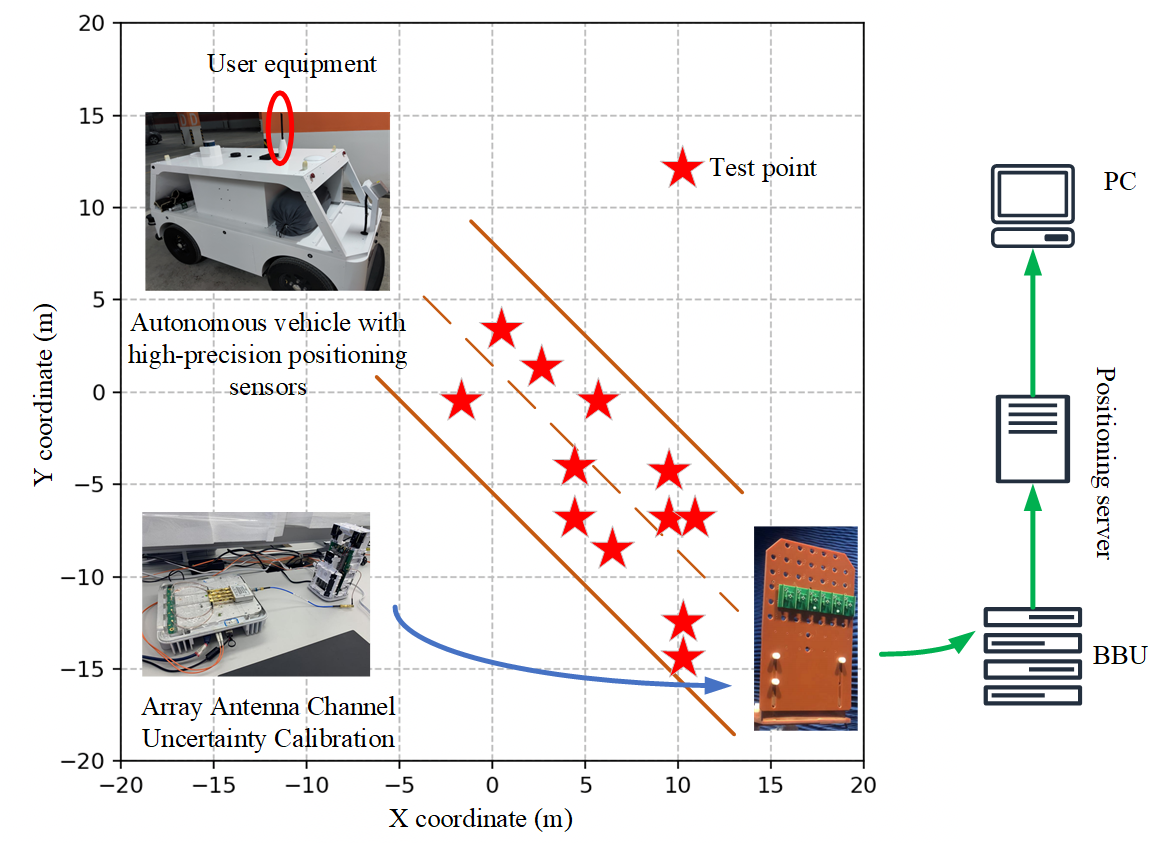}
\vspace{-2.5em}
\caption{Experiment layout for positioning experiments in an underground parking lot.}
\label{p02}
\end{figure}

During the validation stage, the MUSIC  pseudospectrums were used as the criterion for selection. Fig.~\ref{p01} shows the results of the RMSE values of AoA, ToA, and the final positioning error as a function of SNR, with a collection of $500$ data points obtained at each SNR level. Based on the findings obtained from simulation data, it is evident that the performance of the MUSIC algorithm in terms of angle and distance estimation still falls significantly behind intelligent algorithms, even after calibration. In comparison to the CNN, the transformer network exhibits lower RMSE for angle estimation. This outcome validates the superior global feature correlation capability of the Transformer network when processing received signal CSI inputs. Irrespective of whether employing a convolutional network or a transformer network, the approach of parallel training by dividing the angular region demonstrates the potential to enhance the accuracy of angle estimation to a considerable extent. Hence, this methodology effectively augments the fitting capability of the network.

\begin{figure*}[!htpb] \subfloat[]{\includegraphics[width=0.333\textwidth]{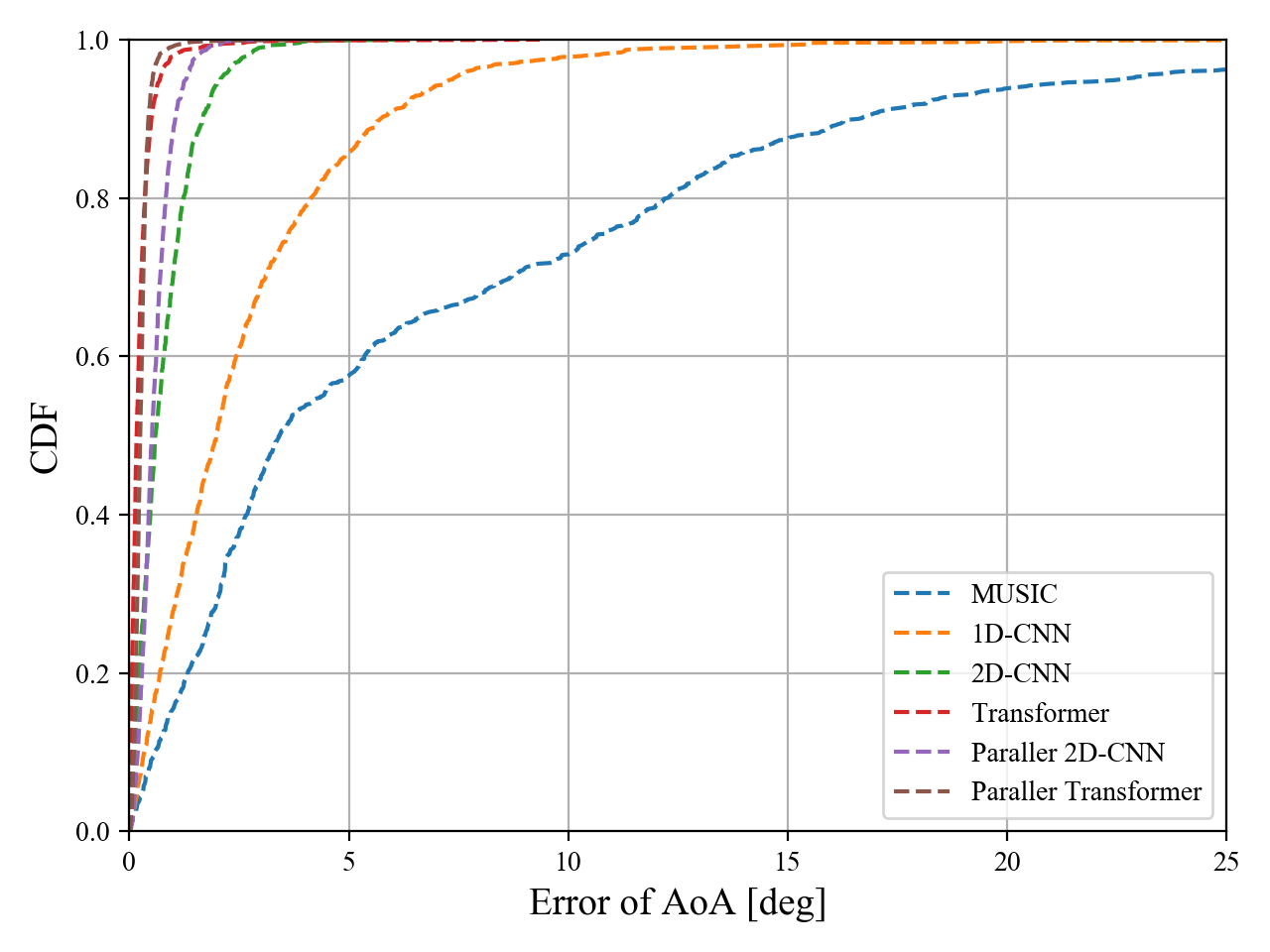}}	\subfloat[]{\includegraphics[width=0.333\textwidth]{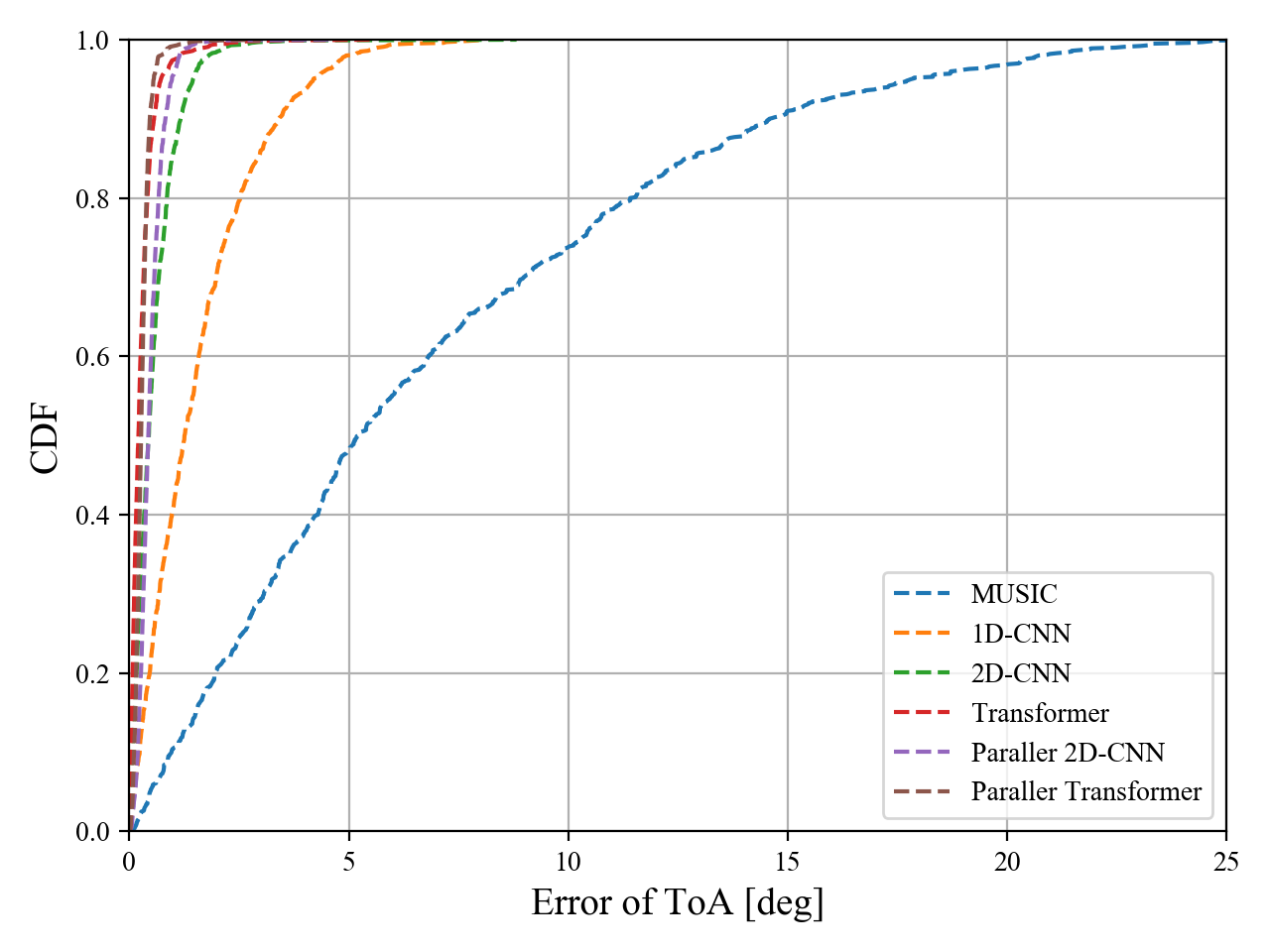}}	\subfloat[]{\includegraphics[width=0.333\textwidth]{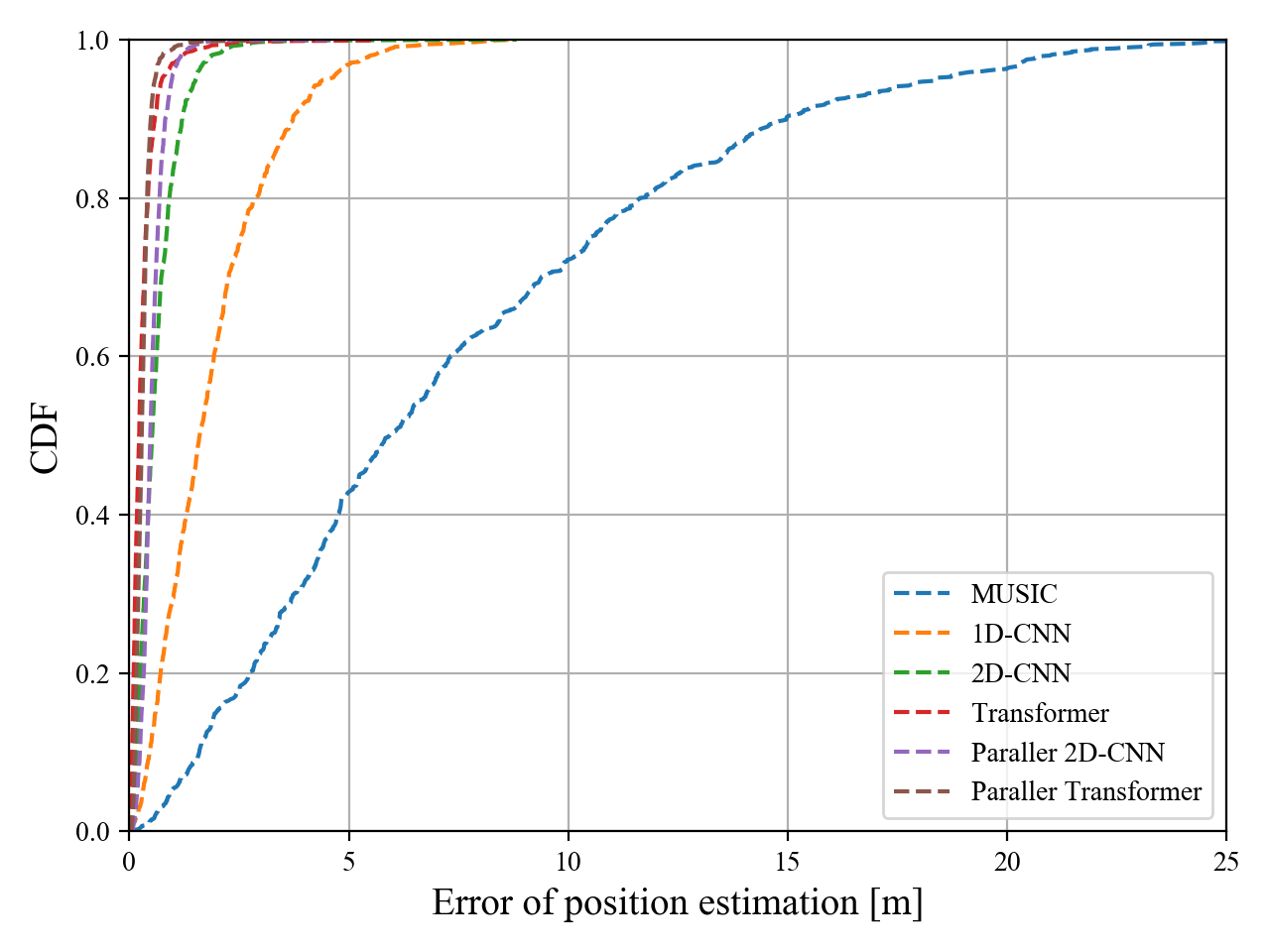}}
\vspace{-1em}
\caption{CDF plots of estimation errors for different algorithms. (a) Error CDF of AoA estimation. (b) Error CDF of ToA estimation. (c) Error CDF of localization estimation. }
\label{p03}
\end{figure*}

\begin{figure}[!htpb]
\centering
\includegraphics[width=0.42\textwidth]{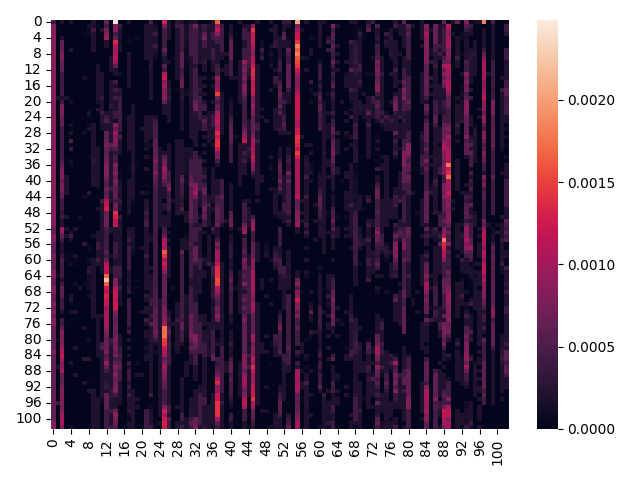}
\vspace{-1.5em}
\caption{Self attention graph of proposed network.}
\vspace{-1em}
\label{p04}
\end{figure}

Furthermore, we examined the angle estimation accuracy achieved by different algorithms within a small angle range $\left[ { - 45^\circ ,45^\circ } \right]$, as given in Fig.~\ref{pic2}. This figure highlights a significant disparity in performance between conventional deep learning algorithms and parallel training methods, particularly within small angle regions. Interestingly, the one-dimensional CNN even yields higher angle measurement errors compared to the MUSIC algorithm. This finding confirms the strong angle dependence of the antenna array gain-phase characteristics, with larger angle regions exhibiting more pronounced phase deviations than smaller angles. The presence of such effect significantly hampers the network's fitting capability, particularly when operating under limited parameters and training data. Consequently, it underscores the need to address this challenge to improve the overall performance of the deep learning models. The results in Fig.~\ref{pic2} also implies that, the MUSIC algorithm can be considered a reliable choice for the network selection process during the validation stage.

In order to further evaluate the efficacy of the proposed method in a practical indoor setting, a field test was conducted within an underground parking lot of Purple Mountain Laboratories utilizing a commercially deployed 5G picocell gNB. The experimental setup, depicted in Fig.~\ref{p02}, involves mounting a 5G UE on an AGV, capable of achieving location accuracy at the sub-meter level. The measured data obtained from this test are analyzed and presented in terms of the error cumulative distribution function (CDF), as illustrated in Fig.~\ref{p03}. The performance observed in the measured data aligns closely with that of the simulated data. From the analysis presented in Fig.~\ref{p03}, it is evident that intelligent algorithms achieve an angle estimation error within $5^\circ$ for $90\%$ of the cases. Similarly, $90\%$ of the distance estimation errors falling within a range of $4\;{\rm m}$. With the exception of the 1D-CNN algorithm, other deep learning algorithms exhibit a positioning error within $3\;{\rm m}$ for $90\%$ of the cases. In contrast, traditional MUSIC algorithms report larger positioning errors, with the errors exceeding $10\;{\rm m}$ for the same percentage of cases. Overall, these findings emphasize the absolute advantage of intelligent algorithms, as they consistently outperform traditional approaches in terms of angle and distance estimation, as well as overall positioning accuracy. Both the transformer network and the parallel training methods demonstrate improvements in the final positioning accuracy, further enhancing the performance of the algorithms.

Finally, to assess the transformer network's capacity to capture global features, we saved the attention matrix $\mathbf{A}$ and subsequently computed the average self-attention matrix. This matrix was then visualized as a heatmap, as displayed in Fig.~\ref{p04}. We readily see from Fig.~\ref{p04} that, the elements proximate to the main diagonal exhibit relatively smaller values, suggesting that the model does not heavily rely on the correlation between the block matrix itself and neighboring blocks. This behavior distinguishes it from conventional transformer and convolutional models. However, in terms of the prediction of angle and distance, our modeling analysis indicates a close relationship with the phase differences across various subcarriers and antenna dimensions. Given the similarity between adjacent blocks, extracting location-related features becomes challenging. Furthermore, the self-attention matrix displays darker, striped patterns in certain columns, signifying the model's emphasis on these blocks' contributions to angle and distance estimation. This capacity for global feature correlation is unavailable in CNNs.

\section{Conclusion}

In this work, we presented a transform-based framework for joint AoA and ToA estimation using monostatic 5G gNB, with existence of hardware impairment. The designed network is advantageous to the baselines in capturing location-related global features, particularly compared to the convolutional networks. We also reveal via anechoic chamber measurement the characteristics of angular-dependent phase error. Then, model-based MUSIC region decision mechanism and corresponding parallel training are incorporated to improve the data consistency. The improved positioning performance of the proposed framework in terms of RMSE and CDF metrics is clearly verified by experiments with synthetic and real datasets.

\begin{acks}
This work was supported in part by the National Natural Science Foundation of China under Grant Nos. 62001103 and U1936201.
\end{acks}

\balance
\bibliographystyle{ACM-Reference-Format}

\end{document}